\documentclass[sigconf,nonacm]{acmart}

\AtBeginDocument{%
  }

\usepackage{multirow}   
\usepackage[table]{xcolor}  
\usepackage{siunitx}    
\usepackage{graphicx}  
\usepackage{booktabs}   
\usepackage{graphicx}   
\usepackage{subcaption}

\begin{document}

\title{PaletteID: Prototype-Composed Semantic Identifiers for Multimodal CTR Prediction}


\author{Huanyu Liu}
\authornote{Equal contribution.}
\affiliation{%
  \institution{Huazhong University of Science and Technology}
  \city{Wuhan}
  \state{Hubei}
  \country{China}
}
\email{huanyuliu001@gmail.com}

\author{Baining Chen}
\authornotemark[1]
\affiliation{%
  \institution{Huazhong University of Science and Technology}
  \city{Wuhan}
  \state{Hubei}
  \country{China}
}
\email{bainingchen4@gmail.com}

\author{Hui Liu}
\authornote{Corresponding author.}
\affiliation{%
  \institution{Huazhong University of Science and Technology}
  \city{Wuhan}
  \state{Hubei}
  \country{China}
}
\email{hliu@hust.edu.cn}

\author{Zengyang Li}
\affiliation{%
  \institution{Central China Normal University}
  \city{Wuhan}
  \state{Hubei}
  \country{China}
}
\email{zengyangli@outlook.com}

\author{Ziyi Huang}
\affiliation{%
  \institution{Huazhong University of Science and Technology}
  \city{Wuhan}
  \state{Hubei}
  \country{China}
}
\email{hzy3406108522@gmail.com}

\renewcommand{\shortauthors}{Huanyu Liu, Baining Chen et al.}

\begin{abstract}
Multimodal information can improve the accuracy of
click-through rate (CTR) prediction and effectively alleviate item cold-start and long-tail problems. Recent studies commonly discretize pretrained multimodal embeddings into semantic
identifiers (SIDs), allowing the model to learn task-specific semantic
representations for recommendation. However, existing methods still provide limited gains due to two major limitations. 
First, codebook assignment fails to preserve semantic relevance and discards fine-grained continuous signals in the original embedding space. 
Second, the residual code paths are highly dependent on prefix codes, which limits the effective representational scalability of hierarchical identifiers. To address these issues, we propose \textit{PaletteID} (PID), a prototype-based semantic identifier. Inspired by palette-based color composition, PID uses a compact set of representative prototype items as semantic anchors to bridge pretrained multimodal content space and recommendation models. 
Specifically, we first construct a prototype palette with Semantic Quality-Aware Determinantal Point Process (SQ-DPP), which jointly considers local content density and global semantic diversity.
Then, for each target item, PID retrieves a sequence of semantically related prototypes and aggregates them into an informative PID representation, enabling rich and complementary semantic modeling.
Extensive experiments on two public datasets demonstrate that PID
consistently improves CTR prediction and yields larger gains for
long-tail items. PID also produces more robust identifier assignments
and provides more interpretable token semantics than existing residual SID
methods.
\end{abstract}



\maketitle

\section{Introduction}
Click-through rate (CTR) prediction is a fundamental component of modern recommender systems, playing a crucial role in traffic allocation and user experience optimization \cite{DBLP:journals/corr/abs-2512-07216, DBLP:journals/corr/abs-2508-01375}. Classical discriminative models typically rely heavily on sparse ID features, which enable the model to memorize frequent user–item interaction patterns. However, such ID-centric paradigms struggle to generalize to long-tail and newly introduced items~\cite{DBLP:journals/corr/abs-2602-11799}. Incorporating intrinsic item content, such as titles, descriptions, images, and videos, therefore offers a promising way to provide more stable and generalizable semantic signals for recommendation~\cite{DBLP:conf/kdd/LiuZYDD0ZZD24}. Recent advances in multimodal foundation models have significantly
improved the quality of multimodal representation learning
~\cite{DBLP:conf/nips/BrownMRSKDNSSAA20, DBLP:conf/icml/RadfordKHRGASAM21}. However, when these
multimodal embeddings are directly fed into CTR models as dense item
features for feature interaction and sequence modeling
~\cite{DBLP:conf/cikm/ShengYGWCZCZG0J24}, the resulting performance
gains often fall short of expectations. In fact, multimodal embeddings
are optimized for general-purpose pretraining objectives, whereas CTR
prediction is driven by user behavior and feedback signals
~\cite{DBLP:conf/cikm/LuoCSYHYLZ0HQZZ25,
DBLP:conf/ecir/YeFSZJ26}. 
To bridge this gap, quantization approaches such as RQ-VAE \cite{DBLP:conf/nips/RajputMSKVHHT0S23} and RQ-KMeans \cite{DBLP:journals/corr/abs-2502-18965} have been proposed to compress multimodal embeddings into compact discrete semantic tokens, namely Semantic IDs (SIDs). SIDs partition the content space into coarse-grained semantic regions,
allowing semantically related items to share recommendation knowledge. Each SID token is mapped to a trainable embedding and incorporated into CTR models as
an item-side feature, enabling its representation to be
optimized with recommendation objectives
~\cite{DBLP:conf/cikm/YeSSWWJ25}.

However, existing SID approaches still face two major limitations that
restrict their effectiveness in CTR prediction. First, hard codebook
assignment discards the continuous structure of multimodal embeddings,
making it difficult to explicitly distinguish different degrees of
semantic relatedness between items. Second, deeper residual codes are prefix-dependent. The same code index
may represent different content semantics under different prefixes,
while sharing its embedding during CTR training mixes inconsistent
collaborative signals. Prefix-conditioned embeddings alleviate this
ambiguity but grow combinatorially with code depth
~\cite{DBLP:conf/cikm/JuCNKW0S25,
DBLP:conf/recsys/ZhengHPRWXNL00L25}.
Consequently, residual SIDs do not scale effectively with code depth,
as deeper levels provide diminishing effective information gains
~\cite{DBLP:journals/corr/abs-2602-05663}. Beyond these performance bottlenecks, for hard semantic identifiers, small perturbations in the multimodal embedding space, such as minor changes to item titles, may cause abrupt identifier changes for items near code-region boundaries, and this instability is amplified at deeper quantization levels. Moreover, abstract codebook indices provide
limited semantic interpretability, making it difficult to trace the
content semantics represented by individual tokens.

To address these issues, we propose PaletteID (PID), a prototype-based identifier that represents each item as a soft composition of semantic prototypes. 
Inspired by the way a limited set of base colors can be blended into rich visual patterns, PID uses a compact prototype palette to bridge pretrained multimodal semantics and behavior-oriented recommendation signals. 
Specifically, we first select a set of representative prototype items from the multimodal item space using Semantic Quality-Aware Determinantal Point Process (SQ-DPP). For each target item, PID retrieves a top-\(K\) sequence of semantically related prototypes as multiple distinct and complementary semantic tokens, providing broad semantic coverage while maintaining high relevance.
We then aggregate their trainable embeddings with similarity-aware weights, thereby preserving fine-grained continuous multimodal proximity.

The contributions are summarized as follows:

\begin{itemize}
    \item We analyze the limitations of SID methods in CTR prediction, including the loss of continuous semantic signals and the limited information gain obtained.
    Accordingly, we propose PID, a prototype-based identifier that preserves multimodal similarity while providing higher compositional capacity.
    
    \item We propose SQ-DPP for prototype palette construction. 
    Unlike reconstruction or clustering methods, we directly select real items as prototype semantic anchors by jointly considering global semantic diversity and local content density. 
    
    \item Experiments on two public datasets show that PID outperforms existing residual SID methods in CTR prediction, long-tail generalization, robustness and interpretability. 
\end{itemize}
\section{Related Work}
\subsection{Content-aware Recommendation}

Content features describe the intrinsic semantics of items independently of observed user interactions and therefore provide valuable signals for cold-start and long-tail recommendation ~\cite{DBLP:journals/iotj/HuangXNZW19, mu2023multimodal}. Incorporating such features reduces the reliance of recommendation models on historical user-item interactions. CB2CF learns a mapping from content features to collaborative filtering embeddings, thereby enriching item representations with auxiliary content semantics ~\cite{DBLP:conf/recsys/BarkanKYK19}. Huang et al. adopt a Siamese architecture to model item content representations and improve item matching in long-tail scenarios through similarity estimation ~\cite{DBLP:conf/kdd/HuangWZX21}. Recently, quantization-based methods have been introduced to compress multimodal representations into learnable semantic IDs for end-to-end training ~\cite{DBLP:conf/cikm/LuoCSYHYLZ0HQZZ25}. Zheng et al. further investigates how different prefix-token parameterization strategies affect recommendation performance ~\cite{DBLP:conf/recsys/ZhengHPRWXNL00L25}. PMMAE designs a multimodal clustering assignment module to encourage consistency and complementarity across different modalities ~\cite{DBLP:conf/cikm/ZhengGYWC25}. STORE decomposes high-cardinality features into a set of stable semantic tokens, aiming to mitigate feature heterogeneity and sparsity ~\cite{DBLP:journals/corr/abs-2511-18805}. Although these methods can capture content similarities among items, our work focuses on preserving fine-grained multimodal semantics when adapting content representations to behavior-driven CTR modeling.
\subsection{Diversity in Recommender Systems}

Items in a recommendation corpus naturally exhibit diverse semantic
characteristics. Existing studies primarily exploit item diversity at the list level to reduce redundancy. Determinantal Point Processes (DPPs) provide a principled probabilistic framework for modeling list-level relevance and diversity ~\cite{macchi1975coincidence}. Industrial DPP-based reranking methods further introduce tunable radial basis function (RBF) kernels to measure semantic distances between items ~\cite{DBLP:conf/cikm/WilhelmRBJCG18}, while personalized DPP extends this idea by adapting the diversity trade-off to individual users ~\cite{DBLP:journals/corr/abs-2004-06390}. Since exact DPP MAP inference is NP-hard, greedy approximation is commonly used in large-scale scenarios; Chen et al. accelerate this process with incremental Cholesky updates ~\cite{DBLP:conf/nips/ChenZZ18}. SSD further computes volume gains via incremental orthogonalization for efficient diversified selection ~\cite{DBLP:conf/kdd/HuangWZX21}. Unlike these list-level diversification methods, our work leverages diversity for item representation construction.
\section{Preliminaries}
\label{sec:preliminary}
\subsection{Task Definition for CTR Prediction}

Click-through rate (CTR) prediction aims to estimate the probability that a user clicks a target item under a given recommendation context. 
It is commonly formulated as a binary classification task. 
Let $\mathcal{U}$ and $\mathcal{I}$ denote the user set and item set, respectively. 
For a user $u\in\mathcal{U}$, we denote the historical interaction sequence as 
$\mathcal{B}_u=[i_1,i_2,\ldots,i_L]$. 
Given a target item $i\in\mathcal{I}$ and other contextual features $\mathbf{o}$, the CTR prediction task can be written as:
\begin{equation}
\hat{y}_{u,i}
=
f_{\Theta}(u,\mathcal{B}_u, i, \mathbf{o}),
\quad
y_{u,i}\in\{0,1\},
\end{equation}
where $f_{\Theta}$ denotes a learnable model consisting of feature interaction modules and sequence modeling modules. For the latter, the target item representation is used to attend to the historical behavior sequence and generate a fixed-length user interest representation. 

Multimodal semantic information can be incorporated as additional item-side features for both the target item and historical items. Thus, the CTR prediction model can be written as:
\begin{equation}
\label{ctr}
\hat{y}_{u,i}
=
\operatorname{CTRModel}
\left(
[
\mathbf{e}^{user}_u,
\tilde{\mathbf{e}}^{item}_i,
\tilde{\mathbf{e}}^{seq}_{u,i},
\mathbf{e}^{context}_{u,i}
]
\right),
\end{equation}
where the tilde notation indicates representations enhanced with multimodal semantic features.

\subsection{Multimodal Semantic Quantization}

Semantic ID (SID) methods convert continuous multimodal item embeddings into discrete token sequences, which can be used as categorical item-side features in CTR models. In the standard VQ-VAE paradigm, a multimodal input is first encoded into a latent vector $\mathbf{z}_i = E(\mathbf{x}_i)$, and then assigned to the nearest codeword in a learnable codebook to obtain a single discrete token. Residual quantization extends this single-level quantization scheme to
a hierarchical multi-level structure, where each level quantizes the
remaining residual information from previous levels to capture finer
semantic details. An item can be represented by a residual
semantic code sequence $
\mathbf{c}_i=[c_{i,1},c_{i,2},\ldots,c_{i,L}],
$
where \(c_{i,l}\) denotes the selected code index at the \(l\)-th
quantization level. RQ-KMeans dispenses with the neural encoder-decoder architecture and performs residual quantization directly  via hierarchical K-means clustering. 

For fusion, embeddings from different quantization levels
are looked up independently and summed. However, as discussed above, this sequential residual structure suffers from an expressive capacity bottleneck. In this paper, inspired by
palette-based color composition, we instead propose a compositional encoding scheme for multimodal semantic representation.

\section{Methodology}
\label{Meth}
\begin{figure}[t]
    \centering
    \includegraphics[width=1.0\linewidth]{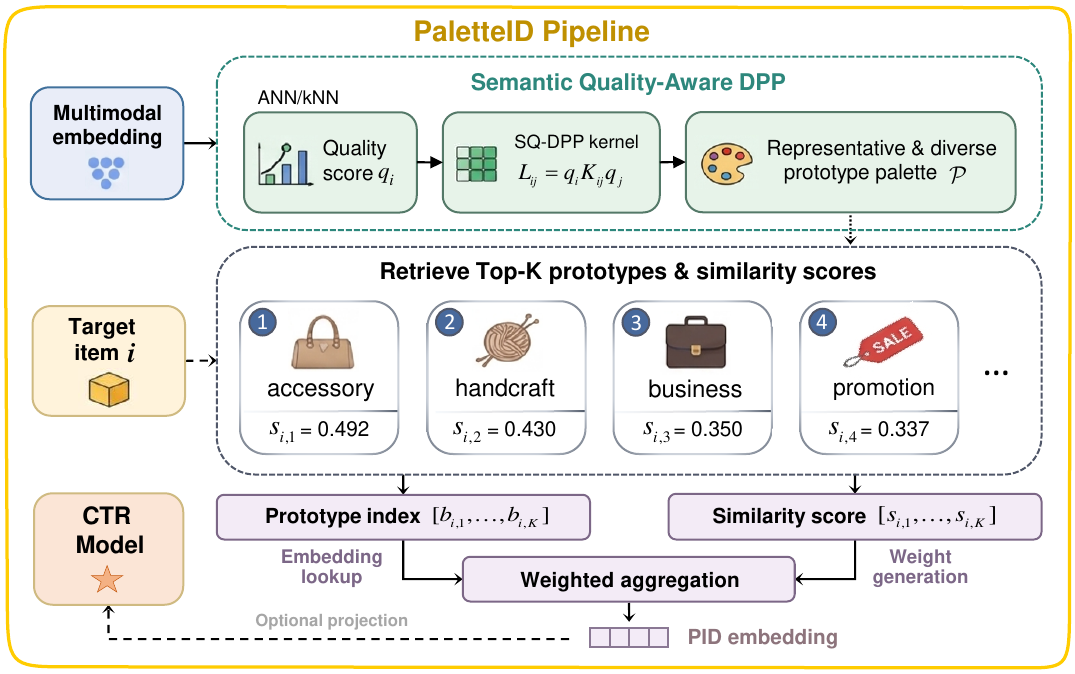}
\caption{Overall architecture of PaletteID. The dashed boxes represent
offline prototype selection and retrieval. During online inference, PID
performs embedding lookup and aggregation to generate
the PID embedding for the CTR model.}
    \label{fig:placeholder}
\end{figure}

Figure~\ref{fig:placeholder} shows the overall pipeline of PID. First, we introduce a Semantic Quality-Aware DPP variant, SQ-DPP, to select a prototype palette from multimodal item embeddings. Second, we retrieve the top-\(K\) most relevant prototypes for each target item and aggregate their trainable embeddings. Finally, the PID embeddings are fed into standard CTR backbones as an additional semantic feature.

\subsection{Prototype Selection}

\subsubsection{Cosine-RBF Kernel}

PID requires an appropriately sized prototype palette to cover the global multimodal item space. However, directly constructing a linear DPP kernel introduces a rank bottleneck, since
\(\mathrm{rank}(\boldsymbol{K}) \leq d\).
Any subset containing more than \(d\) items has a zero determinant. We therefore construct the Cosine-RBF kernel using cosine similarity between multimodal embeddings:
\begin{equation}
K_{ij}
=
\exp\left(
-\gamma
\left(
1-\cos(\hat{\boldsymbol{x}}_i,\hat{\boldsymbol{x}}_j)
\right)
\right),
\end{equation}
where
\(\hat{\boldsymbol{x}}_i
=
\boldsymbol{x}_i/\|\boldsymbol{x}_i\|_2\). \(\gamma\) is the bandwidth governing similarity decay and we use the
median-distance heuristic over randomly sampled item pairs.
This formulation induces an implicit nonlinear feature space, thereby
avoiding the explicit rank ceiling imposed by the embedding dimension.
Up to a rescaling of \(\gamma\), it is equivalent to the standard
radial basis function (RBF) kernel, whose smooth decay captures gradual
semantic redundancy among items \cite{DBLP:conf/cikm/WilhelmRBJCG18}. 

\subsubsection{Semantic Quality-Aware DPP}

To avoid over-selecting isolated items within
low-density regions,
we introduce a content-density-based quality score that guides prototype allocation toward semantically denser and more representative regions.
Specifically, for each candidate item, we perform approximate
nearest-neighbor (ANN) search in the normalized multimodal embedding
space and estimate its content density from the top-\(R\)
retrieved neighbors:
\begin{equation}
\rho_i
=
\sum_{r\in\mathcal{N}_R(i)}
\left[
\frac{
\cos(\hat{\boldsymbol{x}_i},\hat{\boldsymbol{x}_r})-\tau
}{
1-\tau
}
\right]_{+},
\end{equation}
where \(\mathcal{N}_R(i)\) denotes the top-\(R\) neighbors retrieved
by ANN search, \([\cdot]_+=\max(\cdot,0)\), and \(\tau\) is the
similarity threshold. A larger \(\rho_i\) indicates that item \(i\)
lies in a denser semantic region and has stronger local
representativeness. We normalize the density score as
\begin{equation}
q_i
=
\min\left(
\frac{\rho_i}{\tilde{\rho}},
q_{\max}
\right),
\end{equation}
where \(\tilde{\rho}\) is the median density over all candidate items,
and \(q_{\max}\) limits the influence of extremely dense regions.
We construct the SQ-DPP kernel as
\(L_{ij}=q_iK_{ij}q_j\).
Since this corresponds to a diagonal congruence transformation of
the positive semi-definite kernel \(\boldsymbol{K}\),
\(\boldsymbol{L}\) remains positive semi-definite.
For any subset \(S\), its determinant can be decomposed as
\begin{equation}
\det(\boldsymbol{L}_S)
=
\left(\prod_{i\in S}q_i^2\right)
\det(\boldsymbol{K}_S),
\end{equation}
where the multiplicative quality term favors locally representative
prototypes, while \(\det(\boldsymbol{K}_S)\) encourages semantic
complementarity.

\subsubsection{Efficient Global Prototype Selection}

Unlike feed-level re-ranking, SQ-DPP performs a one-time global
prototype selection over the entire candidate item pool. Given a
prototype budget \(M\), we solve the cardinality-constrained DPP
MAP problem:
\begin{equation}
\hat{S}
=
\arg\max_{S\subseteq\mathcal{V},\,|S|=M}
\det(\boldsymbol{L}_S).
\end{equation} Since exact DPP MAP inference is NP-hard, we adopt greedy MAP
inference with incremental Cholesky updates~\cite{DBLP:conf/nips/ChenZZ18}. It constructs the prototype palette once
offline without parameter training or iterative
distribution fitting. At the \(t\)-th selection step, updating the marginal
log-determinant gains of all candidates requires \(O(Nt)\) time.
Therefore, given the precomputed quality scores, selecting \(M\) prototypes has a total update complexity of
$
\sum_{t=1}^{M} O(Nt)
=
O(NM^2).
$
When the \(d\)-dimensional Cosine-RBF similarities between the newly
selected prototype and all candidate items are computed on demand,
an additional \(O(Nd)\) cost is incurred at each step, resulting in
an overall complexity of
$
O(NMd+NM^2)
=
O\bigl(NM(d+M)\bigr).
$ Notably, although prototype selection is performed globally, the
required palette size \(M\) is primarily governed by the semantic
covering complexity of the multimodal embedding space rather than
the raw catalog size. Under a fixed semantic resolution, \(M\) need
not grow linearly with the number of items \(N\). The trade-off among representational capacity, storage cost, and computational cost, controlled by \(M\) is further studied in
Sec.~\ref{sec:Experiments}.

\subsection{PID Embedding Generation}
Let
\(\mathcal{P}=\{p_b\}_{b=1}^{M}\)
denote the selected prototype palette.
Given a target item $i$, prototypes below similarity threshold \(\eta\) are removed to filter noisy matches. We then retrieve its top-\(K\) most similar prototypes according to multimodal similarity. 
This retrieval produces a prototype index sequence, namely the PaletteID of item $i$, together with the corresponding similarity scores:
\begin{equation}
\operatorname{PID}_i = [b_{i,1}, b_{i,2}, \ldots, b_{i,K}], 
\quad
\mathbf{s}_i = [s_{i,1}, s_{i,2}, \ldots, s_{i,K}],
\end{equation}
where $b_{i,j}\in\{1,\ldots,M\}$ is the index of the $j$-th retrieved prototype $p_{b_{i,j}}$, and $s_{i,j}$ denotes its multimodal similarity to item $i$. 
We use a binary mask $\mathbf{m}_i$ to indicate valid positions after padding or truncation, where $m_{i,j}\in\{0,1\}$. Then, we obtain the corresponding item-specific prototype embedding sequence through a trainable embedding lookup:
$
\mathbf{E}_i=[\mathbf{e}_{i,1};\mathbf{e}_{i,2};\ldots;\mathbf{e}_{i,K}],
\quad
\mathbf{e}_{i,j}=\operatorname{Lookup}(b_{i,j})
\in\mathbb{R}^{d_e},
$
where $d_e$ is the embedding dimension. PID composes each target item from multiple complementary
prototypes. We use
independent sigmoid gates so that multiple prototypes can
contribute simultaneously. To calibrate the continuous signals, we apply a lightweight affine transformation to the similarity vector:
\begin{equation}
\boldsymbol{\alpha}_i
=
\mathbf{m}_i
\odot
\sigma(
\mathbf{W}\mathbf{s}_i+\mathbf{b}
),
\qquad
\mathbf{e}^{\mathrm{PID}}_i
=
\boldsymbol{\alpha}_i^{T}\mathbf{E}_i,
\end{equation}
where
\(\mathbf{W}\in\mathbb{R}^{K\times K}\) and
\(\mathbf{b}\in\mathbb{R}^{K}\) are trainable parameters. Since the retrieved prototypes are sorted by similarity, we further introduce a monotonic regularization term to encourage higher-ranked prototypes to receive no smaller confidence weights:
\begin{equation}
\mathcal{L}_{\mathrm{mono}}
=
\frac{1}{N}
\sum_{i=1}^{N}
\sum_{j=1}^{K-1}
m_{i,j}m_{i,j+1}
\max(0, \alpha_{i,j+1}-\alpha_{i,j}).
\end{equation}

\subsection{Downstream CTR Application}
The PID module can be readily integrated into standard CTR
prediction models, such as DeepFM \cite{DBLP:conf/ijcai/GuoTYLH17}, DCNV2 \cite{DBLP:conf/www/WangSCJLHC21}, and DIN \cite{DBLP:conf/kdd/ZhouZSFZMYJLG18}. When required by the downstream backbone, a lightweight nonlinear projection is applied to the aggregated PID embedding to align it with the representation space of other feature embeddings. For both the target item and each item in the historical interaction sequence, the corresponding PID embedding is treated as an additional item-side semantic feature.
For item $i$, the final model output is given by Equation \eqref{ctr}. The model is optimized with binary cross-entropy, and the overall training objective is:
\begin{equation}
\begin{gathered}
\mathcal{L}_{\mathrm{CTR}}
=
-\frac{1}{|\mathcal{D}|}
\sum_{(u,i)\in\mathcal{D}}
\left[
y_{u,i}\log \hat{y}_{u,i}
+
(1-y_{u,i})\log(1-\hat{y}_{u,i})
\right], \\
\mathcal{L}
=
\mathcal{L}_{\mathrm{CTR}}
+ \mathcal{L}_{\mathrm{mono}}.
\end{gathered}
\end{equation}

In deployment, prototype retrieval and multimodal similarity
computation are performed offline in batches. For each item, its
PID sequence and similarity scores are precomputed
and stored in the feature store. During online serving, the model
only retrieves these cached features and performs \(K\) prototype
embedding lookups, a lightweight affine transformation, and weighted
aggregation. Therefore, PID introduces low additional computational
overhead and can be readily integrated into existing serving pipelines.

\section{Experiments}
\label{sec:Experiments}
We conduct extensive experiments to evaluate the effectiveness of PID. 
The results show that PID consistently improves CTR prediction, brings stronger gains on long-tail items, produces more robust semantic assignments and offers clearer semantic interpretability.

\subsection{Experimental Settings}
\label{sec:Experimental Settings}
\subsubsection{Metrics and Datasets}
\label{sec:Datasets}
We use AUC and GAUC to measure ranking model prediction
performance. AUC quantifies the probability of a positive user-item pair outscoring a negative one and GAUC aggregates user-level AUC scores, weighted by the number of impressions for each user. We employ two real-world datasets in our experiments. The key statistics are summarized in Table ~\ref{tab:statistics}.

\textbf{TAOBAO-MM}\footnote{\url{https://taobao-mm.github.io}}.
 An open multimodal dataset derived from real traffic of Taobao’s display advertising system, equipped with item text descriptions and product visual features alongside authentic user click logs. We sample 1.0M unique items from the full item pool and accordingly retain 7.2M valid users. Historical behavior sequences are uniformly truncated to a maximum length of 200.

\textbf{KuaiRec}\footnote{\url{https://kuairec.com}} is a real video recommendation dataset collected from Kuaishou. We define interactions with a watch ratio greater than 3.0 as positive instances indicating strong repeated-viewing engagement, and treat the remaining observed interactions as negative instances. Historical interaction sequences are truncated to 100.

\begin{table}[htbp]
	\centering
	\caption{Dataset statistics}
       \vspace{-4pt} 
	\label{tab:statistics}
	\begin{tabular}{cccc}
		\toprule
		\textbf{Dataset} & \textbf{\#Users} & \textbf{\#Items}& \textbf{\#Instances} \\
		\midrule
		{TAOBAO-MM} & 7.2M &1.0M & 25.9M\\
		{KuaiRec}  & 3.8K & 10.7K & 37.9K \\
		\bottomrule
	\end{tabular}
\end{table}

\subsubsection{Baselines and Backbones}
\label{sec:Compared Models}
To conduct a comprehensive evaluation, we compare the proposed method along three key dimensions: \textbf{(a)} \textbf{Discretization Strategies:} We evaluate the foundational quantization methods introduced in Sec.~\ref{sec:preliminary}, including VQ-VAE \cite{DBLP:conf/nips/OordVK17}, RQ-VAE \cite{DBLP:conf/nips/RajputMSKVHHT0S23}, and a non-parametric clustering variant RQ-KMeans \cite{DBLP:journals/corr/abs-2502-18965}.
\textbf{(b)} \textbf{CTR Backbones:} We employ DCNV2 \cite{DBLP:conf/www/WangSCJLHC21} and RankMixer \cite{DBLP:conf/cikm/ZhuFZJWHDWZGYCC25} to verify performance under distinct feature interaction paradigms. DCNV2 models feature interactions over concatenated field embeddings, whereas RankMixer maintains field-wise token representations before token mixing.
\textbf{(c)} \textbf{Sequence Modeling:} We utilize DIN \cite{DBLP:conf/kdd/ZhouZSFZMYJLG18} with SID/PID as item features for sequential modeling. A “+” suffix on a backbone indicates that the sequential representation from DIN is incorporated as an additional input feature.

\begin{table}[h]
\centering
\caption{Dataset-specific implementation details. }
\label{tab:implementation_details}
\resizebox{\linewidth}{!}{
\begin{tabular}{cccccccc}
\toprule
\textbf{Dataset}
& RQ Structure
& \(R\)
& \(\tau\)
& \(q_{\max}\)
& \(M\)
& \(\eta\)
& \(K\)\\
\midrule
TAOBAO-MM
& \(3\times128\)
& \(50\)
& \(0.5\)
& \(1.8\)
& \(650\) 
& \(0.2\)
& \(12/15\)\\
KuaiRec
& \(2\times64\)
& \(50\)
& \(0.3\)
& \(1.2\)
& \(200\) 
& \(0.2\)
& \(7\)\\
\bottomrule
\end{tabular}
}
\end{table}

\subsubsection{Implementation Details} All methods are implemented in
PyTorch \cite{DBLP:conf/nips/PaszkeGMLBCKLGA19}. The training optimizer is Adam \cite{DBLP:journals/corr/KingmaB14}. \textbf{(a) Semantic identifier methods.} The TAOBAO-MM dataset provides high-quality multimodal features semantically aligned via contrastive learning. For the KuaiRec dataset, we use the BGE-M3 \cite{DBLP:journals/corr/abs-2402-03216} semantic encoder to generate dense text embeddings from item information. The VQ-VAE uses a codebook of size 256. For both RQ-VAE and RQ-KMeans, the structures are presented in
Table~\ref{tab:implementation_details}. For PID, the hyperparameters related
to semantic similarity and prototype coverage are also selected separately
for each dataset based on validation performance. \textbf{(b) CTR backbones.} The batch size is 5120. Learning rate and L2 regularization are tuned
in \{1e-6, 1e-5, 1e-4, 1e-3\}. The embedding dimension is set to 16. The model depth is 2. The dropout is 0.01. For DCNV2, the MLP has hidden layers of (512,512,512). For RankMixer, we set the model hidden dimension to 64. The auto-split flattens input embeddings into 8 tokens on TAOBAO-MM and 16 tokens on KuaiRec, respectively.

 \begin{table}[t]
\centering
\setlength{\tabcolsep}{4.5pt}
\caption{Prediction performance of different semantic identifier methods in terms of AUC and GAUC. A ``+'' suffix denotes DIN sequential features added to the backbone.}
\label{tab:main_ctr}
\begin{tabular}{llcccc}
\toprule
\multirow{2}{*}{\textbf{Backbone}} & \multirow{2}{*}{\textbf{Method}} 
& \multicolumn{2}{c}{\textbf{TAOBAO-MM}} 
& \multicolumn{2}{c}{\textbf{KuaiRec}} \\
\cmidrule(lr){3-4} \cmidrule(lr){5-6}
& & \textbf{AUC} & \textbf{GAUC} & \textbf{AUC} & \textbf{GAUC} \\
\midrule
\multirow{5}{*}{DCNV2$_+$}
& None       & 0.6275 & 0.6242 & 0.8159 & 0.7409 \\
& VQ-VAE     & 0.6284 & 0.6252 & 0.8172 & 0.7426 \\
& RQ-VAE     & 0.6297 & 0.6269 & 0.8173 & 0.7427 \\
& RQ-KMeans  & 0.6295 & 0.6267 & 0.8169 & 0.7422 \\
& PID (Ours)      & \textbf{0.6314} & \textbf{0.6285} & \textbf{0.8185} & \textbf{0.7450} \\
\midrule
\multirow{5}{*}{RankMixer$_+$}
& None       & 0.6286 & 0.6258 & 0.8172 & 0.7416 \\
& VQ-VAE     & 0.6292 & 0.6267 & 0.8178 & 0.7425 \\
& RQ-VAE     & 0.6306 & 0.6285 & 0.8183 & 0.7442 \\
& RQ-KMeans  & 0.6302 & 0.6281 & 0.8180 & 0.7441 \\
& PID (Ours)       & \textbf{0.6318} & \textbf{0.6292} & \textbf{0.8201} & \textbf{0.7514} \\
\bottomrule
\end{tabular}
\end{table}

\subsection{CTR Prediction Performance}
Table~\ref{tab:main_ctr} reports the CTR prediction performance of different semantic identifier methods and ``None'' denotes the backbone without additional content-based semantic identifiers. Three key observations can be made. (1) Incorporating multimodal semantic information generally improves AUC and GAUC over the None baseline, indicating that item content
provides complementary item-side signals for CTR prediction. (2) PID achieves the best performance across all evaluated settings, suggesting the benefit of preserving graded prototype--item similarity during semantic identifier construction. Existing SID methods discretize multimodal embeddings into codebook tokens, where each item is represented by a sequence of hard-assigned codewords. Such hard assignment may discard fine-grained semantic proximity between items. In contrast, PID represents each target item by a combination of multiple semantic prototypes. Since the retrieved prototypes serve as parallel, non-hierarchical semantic anchors, they can be flexibly composed to match different semantic facets of the target item. By using continuous multimodal similarity to calibrate the soft weights, PID provides a more expressive and fine-grained item representation, which better bridges multimodal content semantics and downstream behavior-driven CTR modeling.
(3) PID achieves consistent gains on both DCNV2 and RankMixer, indicating that its gains are not specific to a single feature interaction paradigm.

\subsection{PID vs SID}
In this section, we compare PID with SID from four perspectives. 
First, we examine whether PID facilitates more effective semantic knowledge sharing for long-tail items. Second, we analyze the representation capacity of PID and residual SID. Third, we evaluate the robustness of identifier assignment under perturbations. Finally, we examine the semantic coverage of PID and assess the interpretability of its real-item prototype compositions.
\subsubsection{Long-tail Performance Analysis}

To investigate whether PID can effectively transfer multimodal semantic knowledge to long-tail items, we conduct a bucket-level AUC analysis. 
Following the item-frequency distribution in the training set, we divide items into three groups according to item popularity: Head (20\%), Torso (40\%), and Tail (40\%). As shown in Table~\ref{tab:bucket_auc_taobao}, the gains from multimodal semantic enhancement are relatively modest for Head items, whose item ID embeddings have already been sufficiently optimized through abundant user interactions. 
In contrast, the gains become increasingly pronounced as item popularity decreases. 
These results suggest that shared semantic representations can effectively compensate for sparse behavioral supervision by allowing long-tail items to leverage transferable signals from semantically related items.
\begin{table}[ht]
\centering
\caption{Performance comparison on TAOBAO-MM with DCNV2 across item-popularity groups.}
\label{tab:bucket_auc_taobao}
\begin{tabular}{lcccc}
\toprule
\multirow{2}{*}{\textbf{Method}} & \multicolumn{4}{c}{\textbf{AUC}} \\
\cmidrule(lr){2-5}
& Head (20\%) & Torso (40\%) & Tail (40\%) & Overall \\
\midrule
None       & 0.6335 & 0.5885 & 0.5819 & 0.6275 \\
RQ-VAE     & 0.6352 & 0.5952 & 0.5905 & 0.6297 \\
\midrule
PID        & 0.6366 & 0.5983 & 0.5947 & 0.6314 \\
vs. None   & +0.0031 & +0.0098 & +0.0128 & +0.0041 \\
vs. RQ-VAE & \textbf{+0.0014}& \textbf{+0.0031} & \textbf{+0.0042} & \textbf{+0.0017} \\
\bottomrule
\end{tabular}
\end{table}

As shown in Table~\ref{tab:bucket_auc_taobao}, PID consistently achieves the best performance across all popularity groups, with the largest improvement observed in the Tail group. 
This advantage can be attributed to the different ways in which PID and residual SID methods organize and share semantic information. Because reconstruction objectives are dominated by the empirical item distribution, quantization-based methods such as RQ-VAE may favor densely populated semantic regions, providing less precise coverage of less common semantic patterns. 
In contrast, PID constructs a globally shared palette of representative semantic anchors and composes each item from multiple complementary prototypes. 
Such soft prototype composition enables long-tail items to preserve finer-grained semantic distinctions and access more effective shared behavioral signals, leading to stronger improvements under sparse interaction regimes.

\subsubsection{Information Capacity and Scalability}

For residual SID methods, the maximum number of distinct semantic paths
is determined by the Cartesian product of the codebooks across
quantization levels. In contrast, PID retrieves up to \(K\) distinct
prototypes from a global palette of size \(M\). Ignoring the
similarity-induced ordering, its discrete compositional space contains
up to
\(\sum_{k=1}^{K}\binom{M}{k}\)
prototype combinations. PID further assigns continuous soft gates to
the retrieved prototypes, producing representations within a bounded
continuous region generated by their embeddings.

Although residual SID methods can theoretically expand their code space by introducing additional quantization levels, their effective scalability is constrained by the parameterization of hierarchical code paths. Assigning a distinct embedding to every complete code tuple
$(c_1,c_2,\ldots,c_L)$
can eliminate cross-prefix semantic ambiguity, but requires a number of embeddings that grows exponentially with code depth ~\cite{DBLP:conf/recsys/ZhengHPRWXNL00L25}. Alternatively, sharing a code embedding across prefixes is more
parameter-efficient, but the same deeper-level code may correspond to
different residual semantics under different prefixes. Such sharing
can lead to semantically misaligned parameter updates and potentially
conflicting behavioral gradients. The cross-prefix consistency
analysis in Appendix~\ref{app:code_consistency} further shows that
this ambiguity becomes more pronounced at deeper quantization levels. In contrast, PID flexibly accommodates items of varying semantic complexity. Through similarity-threshold filtering, it allocates a variable number of semantically complementary prototypes to each item from a flat palette, without recursively conditioning each
semantic unit. This decouples the prototype bases and
allows the representation capacity to scale through both the number of
retrieved prototypes and the palette size. We next examine the effects
of the prototype sequence length \(K\) and palette size \(M\):

  \begin{figure}[ht]
	\centering
	\vspace{0.8em}
	\includegraphics[width=1\linewidth]{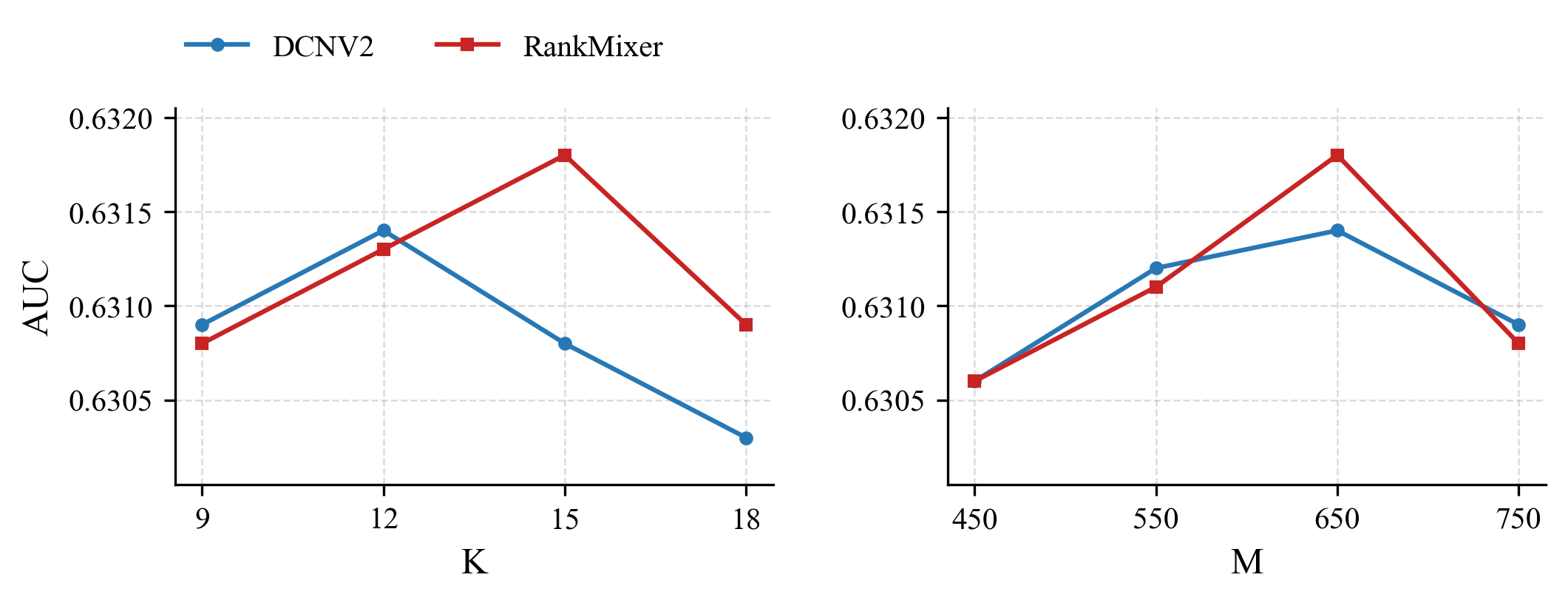} 
	
	\begin{minipage}{0.54\linewidth}
		\centering
		\vspace{-0.8em} 
		\subcaption{Prototype sequence length $K$} \label{fig:length}
	\end{minipage}
	\hspace{-0.025\linewidth}
	\begin{minipage}{0.455\linewidth}
		\centering
		\vspace{-0.8em} 
		\subcaption{Prototype palette size $M$} \label{fig:palette}
	\end{minipage}
	
	\vspace{-0.5em}
	\caption{AUC changes under different prototype sequence length $K$ and prototype palette size $M$ on TAOBAO-MM.}
	\label{fig:hyper_sensitivity}
\end{figure}

 \textbf{Impact of Retrieval Size $K$}. The retrieval size K controls the number of semantic components used to represent each item, balancing semantic richness against the risk of introducing less relevant prototypes. As shown in Figure~\ref{fig:length}, both models exhibit an inverted-U-shaped trend. Increasing K initially improves performance by incorporating broader and more complementary semantic information. However, when K becomes excessively large, lower-similarity prototypes introduce noise and degrade prediction accuracy. Notably, the optimal retrieval size differs between the two backbones, which we attribute to their differing noise sensitivity.
 
  \textbf{Impact of Palette Size $M$}. The palette size $M$ dictates the resolution of the global semantic space. As shown in Fig.~\ref{fig:palette}, both backbones exhibit the same trend. Increasing $M$ enhances the fine-grained representational capacity and semantic coverage of PID; however, an excessively large M may reduce the update frequency of individual basis prototypes and may trigger overfitting. Notably, the optimal palette size increases from \(200\) on KuaiRec to \(650\) on TAOBAO-MM, while the item catalog grows from \(10.7\)K to \(1.0\)M items. This observation suggests that the required palette capacity is primarily governed by the semantic complexity of the item space and grows substantially more slowly than the catalog size. These results provide empirical support for the practical scalability of PID to large item catalogs, as discussed in Sec.~\ref{Meth}.

\subsubsection{Robustness of Identifier Assignment}

We further evaluate the stability of different identifier generation mechanisms under small perturbations in the multimodal input space. Specifically, we add Gaussian noise with different standard deviations to the normalized item embeddings and then regenerate the corresponding SID codes and PID prototype assignments using the frozen RQ-VAE quantizer and the PID prototype palette. To make the comparison between SID and PID fair, we represent both identifiers as assignment vectors. For SID, the residual code path is converted into a concatenation of one-hot vectors $\mathbf{a}_i^{\mathrm{SID}}$. For PID, the top-\(K\) prototype sequence is represented as a sparse weighted vector over the prototype palette $\mathbf{a}_i^{\mathrm{PID}}$. We then compute the assignment cosine between the original and perturbed identifiers:
\[
\mathrm{AssignCos} = \frac{1}{|\mathcal{I}|}\sum_i \cos\big(\mathbf{a}_i^{(0)},\mathbf{a}_i^{(1)}\big).
\]
A higher value indicates a more stable identifier assignment. We also report the change rate of each code level for SID. For PID, we report the weighted assignment change $1-\mathrm{WOverlap}$, where
\[
\mathrm{WOverlap} = \frac{1}{|\mathcal{I}|}\sum_i \sum_{p\in P_i^{(0)}\cap P_i^{(1)}} \min\big(w_{ip}^{(0)},w_{ip}^{(1)}\big),
\]
and $w_{ip}$ denotes the softmax-normalized similarity weight of prototype $p$. This metric is suitable for PID because even if some low-weight prototypes change, the assignment remains stable as long as the major high-weight prototypes are preserved.

\begin{figure}[htb]
	\centering
	\vspace{0.8em}
	\includegraphics[width=1\linewidth]{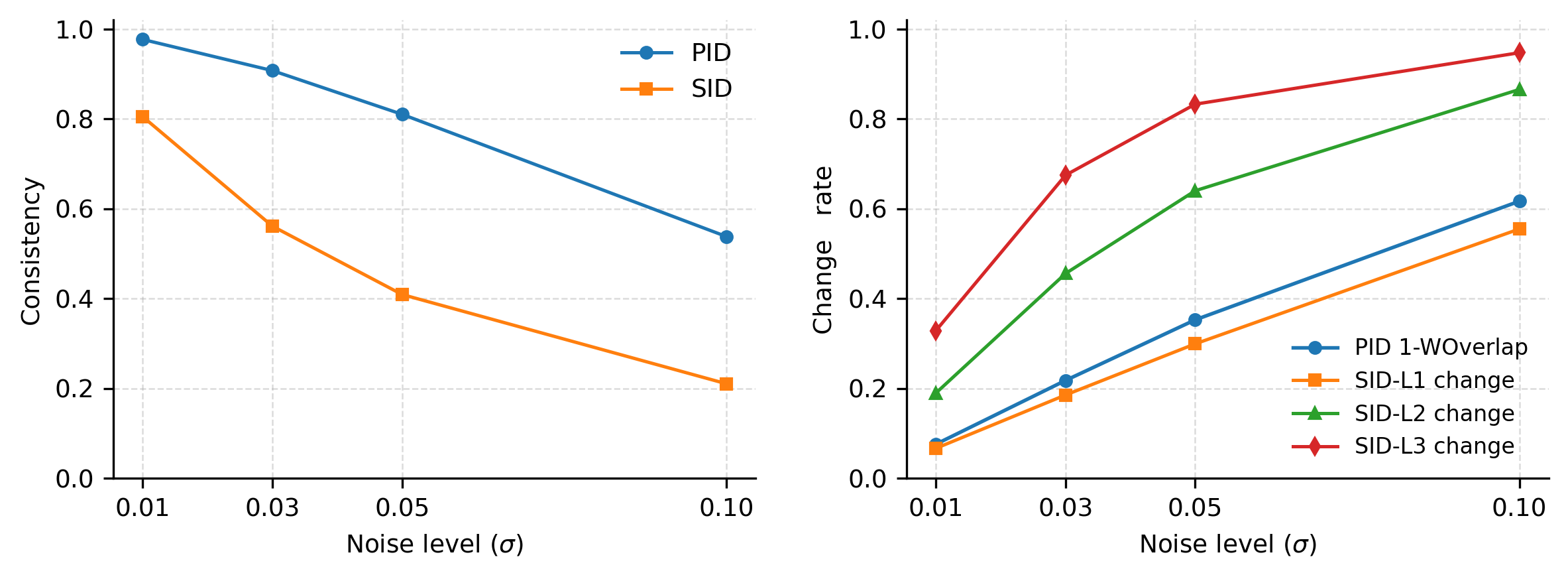} 
	
	\begin{minipage}{0.54\linewidth}
		\centering
		\vspace{-0.8em} 
		\subcaption{Assignment cosine similarity} \label{roba}
	\end{minipage}
	\hspace{-0.025\linewidth}
	\begin{minipage}{0.455\linewidth}
		\centering
		\vspace{-0.8em} 
		\subcaption{Assignment change rate} \label{robb}
	\end{minipage}
	
	\vspace{-0.5em}
    \caption{Identifier assignment stability under perturbations to
    multimodal item embeddings on TAOBAO-MM, comparing PID with RQ-VAE SID.}
	\label{rob}
\end{figure}
\begin{table*}[t]
\centering
\setlength{\tabcolsep}{12pt}  
\caption{A case study of PID prototype composition on KuaiRec. The target video is about handcraft, small-business entrepreneurship, and traditional accessory/object making.}
\label{tab:pid_case}
\begin{tabular}{c c l l}
\toprule
Prototype ID & Sim & Semantic role & Covered target facet \\
\midrule
9077 & 0.4918 & Accessory/product anchor & Object/accessory making, handmade product \\
4807 & 0.4298 & Handcraft/business anchor & Small business, home-based handcraft, side business \\
6361 & 0.3498 & Promotion/commercial anchor & Content promotion, commercial context \\
4845 & 0.3371 & Platform/popularity anchor & High-quality video, popularity signal \\
4812 & 0.3166 & Creator-incentive anchor & Creator incentive, platform promotion \\
\bottomrule
\end{tabular}
\end{table*}
As shown in Fig.~\ref{roba}, PID is consistently more stable than evaluated RQ-SID under embedding perturbations. As the noise level increases, SID assignment consistency drops sharply, while PID assignment cosine similarity decreases more smoothly from 0.977 to 0.538. Moreover, as shown in Fig.~\ref{robb}, deeper SID levels are more sensitive to perturbations, which indicates that changes in the input embedding can be amplified along the residual quantization path. In contrast, PID changes through a weighted prototype mixture. Although PID uses a longer retrieval size ($K$=12/15 on TAOBAO-MM), its weighted change remains comparable to the first-level SID change rate and remains much lower than the deeper-level SID change rates. This demonstrates that PID maintains stable multi-prototype assignments and most of the dominant prototype mass is preserved under moderate perturbations.

\subsubsection{Semantic Coverage and Prototype Interpretability}

To understand the advantage of PID from a semantic perspective, we investigate whether the retrieved prototypes collectively preserve the semantic facets of a target item. For each target video $i$, we collect its caption, cover text, topic tags, and category information, from which we extract a set of semantic facets $F_i$. Let $F_{i,t}$ denote the facet set associated with the $t$-th prototype in the PID sequence of item $i$. We define the Semantic Facet Coverage at rank $r$ for item $i$ as
\begin{equation}
\mathrm{SFC}_i@r =
\frac{1}{|F_i|}
\sum_{f\in F_i}
\mathbb{I}
\left[
\max_{g\in \bigcup_{t=1}^{r} F_{i,t}}
\mathrm{sim}(f,g)
\ge \lambda
\right],
\end{equation}
where a target facet is considered covered if it matches at least one facet from the $r$ retrieved prototypes. We consider two matching protocols. \emph{Strict SFC} uses exact lexical matching after text normalization. \emph{Semantic SFC} computes cosine similarity between BGE-based facet embeddings~\cite{DBLP:journals/corr/abs-2402-03216}, with the matching threshold set to $\lambda=0.75$. We additionally report Random@5, obtained by sampling five prototypes from the same prototype palette, as a control baseline.

\begin{table}[h]
\centering
\caption{Average Semantic Facet Coverage of PID Prototype Composition on KuaiRec.}
\label{tab:sfc_composition}
\begin{tabular}{lcccc}
\toprule
Matching Metric & Random@5 & SFC@1 & SFC@3 & SFC@5 \\
\midrule
Strict SFC   & 0.1691 & 0.2682 & 0.5051 & \textbf{0.6295} \\
Semantic SFC & 0.4136 & 0.5458 & 0.7862 & \textbf{0.8529} \\
\bottomrule
\end{tabular}
\end{table}

As shown in Table~\ref{tab:sfc_composition}, Random@5 achieves non-negligible coverage, likely because short-video metadata contains common category- and platform-level facets. Nevertheless, its coverage remains lower than that of PID with only the nearest prototype, indicating that the top-ranked prototype effectively captures the dominant semantics of the target item. More importantly, incorporating the next four prototypes yields substantial additional coverage, indicating that the retrieved prototypes are not merely redundant nearest neighbors; instead, later prototypes capture target facets that are not covered by the highest-ranked prototype. PID therefore forms a palette-like representation in which multiple semantic anchors jointly cover different aspects of the target content.

We further present a qualitative case study in Table~\ref{tab:pid_case}. The target video involves handcraft production, small-business entrepreneurship, and traditional accessory making. The highest-ranked prototype mainly captures product-level semantics related to handcrafted accessories, while the second prototype complements it with handcraft and home-based business semantics. The third prototype provides an additional commercial and promotional context. The remaining prototypes capture broader platform-popularity and creator-incentive signals, illustrating that PID can include both item-specific semantics and contextual content facets. Unlike SID methods, which represent items using abstract codeword indices whose meanings are learned implicitly and may depend on residual prefixes, PID uses real items as semantic anchors. Each component of a PID representation can therefore be traced back to concrete prototype content, making the composition process directly inspectable at the item level.
\subsection{Ablation}
\subsubsection{Effect of Continuous Prototype Weighting}
To further investigate how continuous prototype-item similarity contributes to PID, we compare three prototype aggregation strategies: mean pooling and two similarity-aware variants that transform the similarity scores using a linear layer followed by either softmax or sigmoid.

\begin{table}[h]
\centering
\setlength{\tabcolsep}{10pt} 
\caption{Effect of different prototype aggregation strategies on TAOBAO-MM with RankMixer.}
\label{tab:aggregation_ablation}
\begin{tabular}{lccc}
\toprule
\textbf{Aggregation} & \textbf{AUC} & \textbf{GAUC} & \textbf{Tail AUC} \\
\midrule
Mean Pooling   & 0.6310 & 0.6285 & 0.5937 \\
Linear + Softmax & 0.6313 & 0.6286 & 0.5939 \\
Linear + Sigmoid & \textbf{0.6318} & \textbf{0.6292} & \textbf{0.5946} \\
\bottomrule
\end{tabular}
\end{table}

 As shown in Table~\ref{tab:aggregation_ablation}, mean pooling yields the lowest performance, suggesting that assigning equal importance to all retrieved prototypes fails to exploit their different degrees of relevance to the target item. Linear + Softmax improves over mean pooling but remains inferior to Linear + Sigmoid. This result suggests that enforcing unit-sum competition among the retrieved prototypes is less suitable for prototype composition. The prototypes are not mutually exclusive candidates; instead, multiple prototypes may simultaneously capture complementary semantic facets of the same item.  Sigmoid gating assigns an independent confidence weight to each prototype, allowing multiple highly relevant prototypes to contribute simultaneously. Unlike SID methods that represent items through hard-assigned discrete codes, PID combines discrete prototype identities with similarity-derived continuous gates. This design preserves graded prototype-item relevance and enables more fine-grained semantic representations for CTR prediction.

\subsubsection{Effect of Prototype Selection Strategies}
To investigate how prototype palette construction affects PID, we compare different selection strategies using both palette-quality metrics and CTR performance. \textit{AvgLen} denotes the average number of valid prototypes retained for each item after similarity-threshold filtering. \textit{Mean Sim} denotes the average similarity between target items and their retrieved prototypes. \textit{Usage Gini} measures the imbalance in prototype retrieval frequency, with a lower value indicating more even utilization across the palette. For a fair comparison, all methods use the same prototype budget, retrieval size, and similarity threshold.

\begin{table}[h]
\centering
\caption{Effect of prototype selection strategies on TAOBAO-MM with RankMixer.}
\label{tab:prototype_selection}
\begin{tabular}{lcccc}
\toprule
Method  & AvgLen & Mean Sim & Usage Gini & \textbf{AUC} \\
\midrule
Random  & 14.4& 0.280 & 0.19  & 0.6311  \\
KMeans & 14.4 & 0.300&0.17& 0.6313  \\
RBF-DPP & 13.9 & 0.227 &0.23& 0.6312  \\
SQ-DPP & 14.3 & 0.287&0.18 & \textbf{0.6318}  \\
\bottomrule
\end{tabular}
\end{table}

As shown in Table~\ref{tab:prototype_selection}, random sampling can provide non-trivial coverage when the prototype budget is sufficiently large, but it lacks a stable semantic structure. KMeans achieves strong AvgLen and Mean Sim. However, its objective mainly minimizes clustering distortion, which does not necessarily produce the best prototype palette for PID. Pure RBF-DPP promotes global semantic diversity through determinant-based subset selection. Without an explicit representativeness term, however, it may allocate part of the fixed prototype budget to isolated or low-density regions, resulting in lower AvgLen and Mean Sim. SQ-DPP incorporates local content density as a quality signal while retaining the diversity objective of DPP. By combining these two signals, it achieves a favorable balance among prototype relevance, utilization, and semantic complementarity, leading to the best downstream AUC.

\subsubsection{Impact of the Density-Aware Quality Score}

To examine how the density-aware quality score affects prototype
selection, we vary the similarity threshold \(\tau\) and the quality
clipping bound \(q_{\max}\). Coverage@10 denotes the proportion of
items associated with at least ten valid retrieved prototypes after
similarity-threshold filtering, reflecting the depth of prototype
coverage. Mean Top-\(K\) Similarity (\(\mathrm{Sim}_K\)) is the average
cosine similarity between each item and its valid top-\(K\) prototypes,
measuring the semantic relevance of the retrieved prototypes. We
additionally report the clipping rate \(r_{\mathrm{clip}}\), defined as
the fraction of candidate items whose unclipped quality scores exceed
\(q_{\max}\). When varying \(\tau\), we fix \(q_{\max}=1.8\); when
varying \(q_{\max}\), we fix \(\tau=0.5\).
\begin{table}[ht]
\centering
\caption{Impact of similarity threshold $\tau$ and quality clipping bound $q_{max}$ on prototype representation on TAOBAO-MM.}
\label{tab:q_score_analysis}
\resizebox{\columnwidth}{!}{
\begin{tabular}{ccccccccc}
\toprule
Setting & $\overline{\mathcal{N}}$ &  \(\tilde{\rho}\) & $q_{\mathrm{std}}$ & \(r_{\mathrm{clip}}\) & $q_{\mathrm{mean}}$ & \(\mathrm{Sim}_K\) & \(\mathrm{Cov}@10\) & \textbf{AUC} \\
\midrule
$\tau=0.4$ & 50.00 & 15.80 & 0.177 & 0.01 & 1.12 & 0.287 & 0.906 & 0.6314 \\
$\tau=0.5$ & 46.34 & 8.96  & 0.291 & 0.10 & 1.22 &  0.286 & 0.970 & \textbf{0.6318} \\
$\tau=0.6$ & 19.79 & 1.84  & 0.368 & 0.33 & 1.32 & 0.260 & 0.979 & 0.6313 \\
\midrule
$q_{max}=1.5$ & - & - & 0.211 & 0.212 & 1.17 & 0.265 & 0.975 & 0.6314 \\
$q_{max}=1.8$ & - & - & 0.291 & 0.105 & 1.22 & 0.286 & 0.970 & \textbf{0.6318} \\ 
$q_{max}=2.1$ & - & - & 0.339 & 0.045 & 1.24 & 0.273 & 0.960 & 0.6313 \\ 
\bottomrule
\end{tabular}
}
\end{table}

As shown in Table~\ref{tab:q_score_analysis}, increasing \(\tau\)
reduces both the average effective neighborhood size and the median
density. With a small threshold \(\tau=0.4\), even moderately similar
neighbors contribute to the density score. This makes the quality term
favor broad and frequently occurring semantic regions, causing more
prototypes to be selected around these common modes. Consequently, the
retrieved prototypes achieve high prototype--item similarity but provide
shallower coverage of the overall item space. In contrast, a large \(\tau\) produces sparse, high-variance density estimates that are overly sensitive to a few close neighbors, increasing Coverage@10 but reducing \(\mathrm{Sim}_K\). The intermediate setting \(\tau=0.5\) achieves the best balance between prototype relevance and coverage. Similarly, a small \(q_{\max}\) excessively clips high-quality candidates and weakens the preference for locally representative prototypes, whereas an overly large \(q_{\max}\) allows dense semantic regions to exert disproportionate influence and reduces coverage depth. Together, \(\tau=0.5\)  and \(q_{\max}=1.8\) provide the best balance between prototype relevance, coverage depth, and downstream AUC.

\section{Conclusion}

In this work, we propose PaletteID, a prototype-composed semantic
identifier that represents each item through a similarity-aware
combination of real-item prototypes. PID integrates quality- and
diversity-aware prototype selection with continuous semantic
weighting, improving the effective representational capacity,
assignment robustness, and interpretability of semantic identifiers.
Extensive experiments on two public datasets demonstrate consistent
improvements in overall CTR prediction and long-tail item performance. PID also supports efficient deployment. Prototype selection,
similarity retrieval, and PID sequence construction are performed
offline, while online serving requires only prototype embedding lookup,
lightweight weight generation, and aggregation. Future work will
explore personalized prototype palettes and extend PID to retrieval over long user behavior sequences.
\bibliographystyle{ACM-Reference-Format}
\bibliography{sample}

@article{DBLP:journals/corr/abs-2512-07216,
  author       = {Bin Wu and
                  Feifan Yang and
                  Zhangming Chan and
                  Yu{-}Ran Gu and
                  Jiawei Feng and
                  Chao Yi and
                  Xiang{-}Rong Sheng and
                  Han Zhu and
                  Jian Xu and
                  Mang Ye and
                  Bo Zheng},
  title        = {{MUSE:} {A} Simple Yet Effective Multimodal Search-Based Framework
                  for Lifelong User Interest Modeling},
  journal      = {CoRR},
  volume       = {abs/2512.07216},
  year         = {2025},
  url          = {https://doi.org/10.48550/arXiv.2512.07216}
}

@article{DBLP:journals/corr/abs-2508-01375,
  author       = {Yining Yao and
                  Ziwei Li and
                  Shuwen Xiao and
                  Boya Du and
                  Jialin Zhu and
                  Junjun Zheng and
                  Xiangheng Kong and
                  Yuning Jiang},
  title        = {SaviorRec: Semantic-Behavior Alignment for Cold-Start Recommendation},
  journal      = {CoRR},
  volume       = {abs/2508.01375},
  year         = {2025},
  url          = {https://doi.org/10.48550/arXiv.2508.01375}
}

@article{DBLP:journals/corr/abs-2602-11799,
  author       = {Pingjun Pan and
                  Tingting Zhou and
                  Peiyao Lu and
                  Tingting Fei and
                  Hongxiang Chen and
                  Chuanjiang Luo},
  title        = {Hi-SAM: {A} Hierarchical Structure-Aware Multi-modal Framework for
                  Large-Scale Recommendation},
  journal      = {CoRR},
  volume       = {abs/2602.11799},
  year         = {2026},
  url          = {https://doi.org/10.48550/arXiv.2602.11799}
}

@inproceedings{DBLP:conf/kdd/LiuZYDD0ZZD24,
  author       = {Qijiong Liu and
                  Jieming Zhu and
                  Yanting Yang and
                  Quanyu Dai and
                  Zhaocheng Du and
                  Xiao{-}Ming Wu and
                  Zhou Zhao and
                  Rui Zhang and
                  Zhenhua Dong},

  title        = {Multimodal Pretraining, Adaptation, and Generation for Recommendation:
                  {A} Survey},
  booktitle    = {{ACM} {SIGKDD} Conference on Knowledge Discovery
                  and Data Mining ({KDD})},
  pages        = {6566--6576},
  year         = {2024},
  url          = {https://doi.org/10.1145/3637528.3671473},
}

@inproceedings{DBLP:conf/nips/BrownMRSKDNSSAA20,
  author       = {Tom B. Brown and
                  Benjamin Mann and
                  Nick Ryder and
                  Melanie Subbiah and
                  Jared Kaplan and
                  Prafulla Dhariwal and
                  Arvind Neelakantan and
                  Pranav Shyam and
                  Girish Sastry and
                  Amanda Askell and
                  Sandhini Agarwal and
                  Ariel Herbert{-}Voss and
                  Gretchen Krueger and
                  Tom Henighan and
                  Rewon Child and
                  Aditya Ramesh and
                  Daniel M. Ziegler and
                  Jeffrey Wu and
                  Clemens Winter and
                  Christopher Hesse and
                  Mark Chen and
                  Eric Sigler and
                  Mateusz Litwin and
                  Scott Gray and
                  Benjamin Chess and
                  Jack Clark and
                  Christopher Berner and
                  Sam McCandlish and
                  Alec Radford and
                  Ilya Sutskever and
                  Dario Amodei},
  title        = {Language Models are Few-Shot Learners},
  booktitle    = {Advances in Neural Information Processing Systems (NeurIPS)},
  year         = {2020},
  url          = {https://proceedings.neurips.cc/paper/2020/hash/1457c0d6bfcb4967418bfb8ac142f64a-Abstract.html}
}

@inproceedings{DBLP:conf/cikm/ShengYGWCZCZG0J24,
  author       = {Xiang{-}Rong Sheng and
                  Feifan Yang and
                  Litong Gong and
                  Biao Wang and
                  Zhangming Chan and
                  Yujing Zhang and
                  Yueyao Cheng and
                  Yong{-}Nan Zhu and
                  Tiezheng Ge and
                  Han Zhu and
                  Yuning Jiang and
                  Jian Xu and
                  Bo Zheng},
  title        = {Enhancing Taobao Display Advertising with Multimodal Representations:
                  Challenges, Approaches and Insights},
  booktitle    = {{ACM} International Conference on Information
                  and Knowledge Management ({CIKM})},
  pages        = {4858--4865},
  year         = {2024},
  url          = {https://doi.org/10.1145/3627673.3680068}
}

@inproceedings{DBLP:conf/cikm/LuoCSYHYLZ0HQZZ25,
  author       = {Xinchen Luo and
                  Jiangxia Cao and
                  Tianyu Sun and
                  Jinkai Yu and
                  Rui Huang and
                  Wei Yuan and
                  Hezheng Lin and
                  Yichen Zheng and
                  Shiyao Wang and
                  Qigen Hu and
                  Changqing Qiu and
                  Jiaqi Zhang and
                  Xu Zhang and
                  Zhiheng Yan and
                  Jingming Zhang and
                  Simin Zhang and
                  Mingxing Wen and
                  Zhaojie Liu and
                  Guorui Zhou},

  title        = {{QARM:} Quantitative Alignment Multi-Modal Recommendation at Kuaishou},
  booktitle    = {{ACM} International Conference on Information
                  and Knowledge Management ({CIKM})},
  pages        = {5915--5922},
  year         = {2025},
  url          = {https://doi.org/10.1145/3746252.3761502}
}

@inproceedings{DBLP:conf/ecir/YeFSZJ26,
  author       = {Yu Ye and
                  Junchen Fu and
                  Yu Song and
                  Kaiwen Zheng and
                  Joemon M. Jose},
  title        = {Are Multimodal Embeddings Truly Beneficial for Recommendation? {A}
                  Deep Dive into Whole vs. Individual Modalities},
  booktitle    = {European Conference on Information
                  Retrieval ({ECIR})},
  pages        = {66--81},
  year         = {2026},
  url          = {https://doi.org/10.1007/978-3-032-21324-2\_5}
}

@inproceedings{DBLP:conf/nips/RajputMSKVHHT0S23,
  author       = {Shashank Rajput and
                  Nikhil Mehta and
                  Anima Singh and
                  Raghunandan Hulikal Keshavan and
                  Trung Vu and
                  Lukasz Heldt and
                  Lichan Hong and
                  Yi Tay and
                  Vinh Q. Tran and
                  Jonah Samost and
                  Maciej Kula and
                  Ed H. Chi and
                  Mahesh Sathiamoorthy},
  title        = {Recommender Systems with Generative Retrieval},
  booktitle    = {Advances in Neural Information Processing Systems (NeurIPS)},
  year         = {2023},
  url          = {http://papers.nips.cc/paper\_files/paper/2023/hash/20dcab0f14046a5c6b02b61da9f13229-Abstract-Conference.html}
}

@article{DBLP:journals/corr/abs-2502-18965,
  author       = {Jiaxin Deng and
                  Shiyao Wang and
                  Kuo Cai and
                  Lejian Ren and
                  Qigen Hu and
                  Weifeng Ding and
                  Qiang Luo and
                  Guorui Zhou},
  title        = {OneRec: Unifying Retrieve and Rank with Generative Recommender and
                  Iterative Preference Alignment},
  journal      = {CoRR},
  volume       = {abs/2502.18965},
  year         = {2025},
  url          = {https://doi.org/10.48550/arXiv.2502.18965}
}

@inproceedings{DBLP:conf/cikm/YeSSWWJ25,
  author       = {Wencai Ye and
                  Mingjie Sun and
                  Shaoyun Shi and
                  Peng Wang and
                  Wenjin Wu and
                  Peng Jiang},

  title        = {{DAS:} Dual-Aligned Semantic IDs Empowered Industrial Recommender
                  System},
  booktitle    = {{ACM} International Conference on Information
                  and Knowledge Management ({CIKM})},
  pages        = {6217--6224},
  year         = {2025},
  url          = {https://doi.org/10.1145/3746252.3761529}
}

@article{DBLP:journals/corr/abs-2511-18805,
  author       = {Yi Xu and
                  Chaofan Fan and
                  Jinxin Hu and
                  Yu Zhang and
                  Xiaoyi Zeng and
                  Jing Zhang},
  title        = {{STORE:} Semantic Tokenization, Orthogonal Rotation and Efficient
                  Attention for Scaling Up Ranking Models},
  journal      = {CoRR},
  volume       = {abs/2511.18805},
  year         = {2025},
  url          = {https://doi.org/10.48550/arXiv.2511.18805}
}

@inproceedings{DBLP:conf/cikm/ZhengGYWC25,
  author       = {Jiawei Zheng and
                  Hao Gu and
                  Lingling Yi and
                  Jie Wen and
                  Chuan Chen},

  title        = {Personalized Multi Modal Alignment Encoding for CTR-Recommendation
                  in WeChat},
  booktitle    = {{ACM} International Conference on Information
                  and Knowledge Management ({CIKM})},
  pages        = {6301--6308},
  year         = {2025},
  url          = {https://doi.org/10.1145/3746252.3761525}
}

@inproceedings{DBLP:conf/cikm/JuCNKW0S25,
  author       = {Clark Mingxuan Ju and
                  Liam Collins and
                  Leonardo Neves and
                  Bhuvesh Kumar and
                  Louis Yufeng Wang and
                  Tong Zhao and
                  Neil Shah},
  title        = {Generative Recommendation with Semantic IDs: {A} Practitioner's Handbook},
  booktitle    = {{ACM} International Conference on Information
                  and Knowledge Management ({CIKM})},
  pages        = {6420--6425},
  year         = {2025},
  url          = {https://doi.org/10.1145/3746252.3761612}
}

@inproceedings{DBLP:conf/recsys/ZhengHPRWXNL00L25,
  author       = {Carolina Zheng and
                  Minhui Huang and
                  Dmitrii Pedchenko and
                  Kaushik Rangadurai and
                  Siyu Wang and
                  Fan Xia and
                  Gaby Nahum and
                  Jie Lei and
                  Yang Yang and
                  Tao Liu and
                  Zutian Luo and
                  Xiaohan Wei and
                  Dinesh Ramasamy and
                  Jiyan Yang and
                  Yiping Han and
                  Lin Yang and
                  Hangjun Xu and
                  Rong Jin and
                  Shuang Yang},
  title        = {Enhancing Embedding Representation Stability in Recommendation Systems
                  with Semantic {ID}},
  booktitle    = {{ACM} Conference on Recommender Systems,
                  (RecSys)},
  pages        = {954--957},
  year         = {2025},
  url          = {https://doi.org/10.1145/3705328.3748123}
}

@article{DBLP:journals/corr/abs-2602-05663,
  author       = {Shiteng Cao and
                  Junda She and
                  Ji Liu and
                  Bing Zeng and
                  Chengcheng Guo and
                  Kuo Cai and
                  Qiang Luo and
                  Ruiming Tang and
                  Han Li and
                  Kun Gai and
                  Zhiheng Li and
                  Cheng Yang},
  title        = {{GLASS:} {A} Generative Recommender for Long-sequence Modeling via
                  SID-Tier and Semantic Search},
  journal      = {CoRR},
  volume       = {abs/2602.05663},
  year         = {2026},
  url          = {https://doi.org/10.48550/arXiv.2602.05663}
}

@article{macchi1975coincidence,
  title={The coincidence approach to stochastic point processes},
  author={Macchi, Odile},
  journal={Advances in Applied Probability},
  volume={7},
  number={1},
  pages={83--122},
  year={1975}
}

@inproceedings{DBLP:conf/cikm/WilhelmRBJCG18,
  author       = {Mark Wilhelm and
                  Ajith Ramanathan and
                  Alexander Bonomo and
                  Sagar Jain and
                  Ed H. Chi and
                  Jennifer Gillenwater},
  title        = {Practical Diversified Recommendations on YouTube with Determinantal
                  Point Processes},
  booktitle    = {{ACM} International Conference on Information
                  and Knowledge Management ({CIKM})},
  pages        = {2165--2173},
  year         = {2018},
  url          = {https://doi.org/10.1145/3269206.3272018}
}

@article{DBLP:journals/corr/abs-2004-06390,
  author       = {Yichao Wang and
                  Xiangyu Zhang and
                  Zhirong Liu and
                  Zhenhua Dong and
                  Xinhua Feng and
                  Ruiming Tang and
                  Xiuqiang He},
  title        = {Personalized Re-ranking for Improving Diversity in Live Recommender
                  Systems},
  journal      = {CoRR},
  volume       = {abs/2004.06390},
  year         = {2020},
  url          = {https://arxiv.org/abs/2004.06390}
}

@inproceedings{DBLP:conf/nips/ChenZZ18,
  author       = {Laming Chen and
                  Guoxin Zhang and
                  Eric Zhou},
  title        = {Fast Greedy {MAP} Inference for Determinantal Point Process to Improve
                  Recommendation Diversity},
  booktitle    = {Advances in Neural Information Processing Systems (NeurIPS)},
  pages        = {5627--5638},
  year         = {2018},
  url          = {https://proceedings.neurips.cc/paper/2018/hash/dbbf603ff0e99629dda5d75b6f75f966-Abstract.html}
}

@inproceedings{DBLP:conf/kdd/HuangWZX21,
  author       = {Yanhua Huang and
                  Weikun Wang and
                  Lei Zhang and
                  Ruiwen Xu},

  title        = {Sliding Spectrum Decomposition for Diversified Recommendation},
  booktitle    = {{ACM} {SIGKDD} Conference on Knowledge Discovery
                  and Data Mining ({KDD})},
  pages        = {3041--3049},
  year         = {2021},
  url          = {https://doi.org/10.1145/3447548.3467108}
}

@inproceedings{DBLP:conf/recsys/BarkanKYK19,
  author       = {Oren Barkan and
                  Noam Koenigstein and
                  Eylon Yogev and
                  Ori Katz},
  title        = {{CB2CF:} a neural multiview content-to-collaborative filtering model
                  for completely cold item recommendations},
  booktitle    = {{ACM} Conference on Recommender Systems (RecSys)},
  pages        = {228--236},
  year         = {2019},
  url          = {https://doi.org/10.1145/3298689.3347038}
}

@article{DBLP:journals/iotj/HuangXNZW19,
  author       = {Zhenhua Huang and
                  Xin Xu and
                  Juan Ni and
                  Honghao Zhu and
                  Cheng Wang},
  title        = {Multimodal Representation Learning for Recommendation in Internet
                  of Things},
  journal      = {{IEEE} Internet Things J.},
  volume       = {6},
  number       = {6},
  pages        = {10675--10685},
  year         = {2019},
  url          = {https://doi.org/10.1109/JIOT.2019.2940709},
}

@article{mu2023multimodal,
  title={Multimodal movie recommendation system using deep learning},
  author={Mu, Yongheng and Wu, Yun},
  journal={Mathematics},
  volume={11},
  number={4},
  pages={895},
  year={2023}
}

@inproceedings{DBLP:conf/kdd/ZhouZSFZMYJLG18,
  author       = {Guorui Zhou and
                  Xiaoqiang Zhu and
                  Chengru Song and
                  Ying Fan and
                  Han Zhu and
                  Xiao Ma and
                  Yanghui Yan and
                  Junqi Jin and
                  Han Li and
                  Kun Gai},
  title        = {Deep Interest Network for Click-Through Rate Prediction},
  booktitle    = {{ACM} {SIGKDD} International Conference on
                  Knowledge Discovery and Data Mining ({KDD})},
  pages        = {1059--1068},
  year         = {2018},
  url          = {https://doi.org/10.1145/3219819.3219823}
}

@inproceedings{DBLP:conf/nips/OordVK17,
  author       = {A{\"{a}}ron van den Oord and
                  Oriol Vinyals and
                  Koray Kavukcuoglu},
  title        = {Neural Discrete Representation Learning},
  booktitle    = {Advances in Neural Information Processing Systems (NeurIPS)},
  pages        = {6306--6315},
  year         = {2017},
  url          = {https://proceedings.neurips.cc/paper/2017/hash/7a98af17e63a0ac09ce2e96d03992fbc-Abstract.html}
}

@inproceedings{DBLP:conf/cikm/ZhuFZJWHDWZGYCC25,
  author       = {Jie Zhu and
                  Zhifang Fan and
                  Xiaoxie Zhu and
                  Yuchen Jiang and
                  Hangyu Wang and
                  Xintian Han and
                  Haoran Ding and
                  Xinmin Wang and
                  Wenlin Zhao and
                  Zhen Gong and
                  Huizhi Yang and
                  Zheng Chai and
                  Zhe Chen and
                  Yuchao Zheng and
                  Qiwei Chen and
                  Feng Zhang and
                  Xun Zhou and
                  Peng Xu and
                  Xiao Yang and
                  Di Wu and
                  Zuotao Liu},
  title        = {RankMixer: Scaling Up Ranking Models in Industrial Recommenders},
  booktitle    = {{ACM} International Conference on Information
                  and Knowledge Management ({CIKM})},
  pages        = {6309--6316},
  year         = {2025},
  url          = {https://doi.org/10.1145/3746252.3761507}
}

@inproceedings{DBLP:conf/www/WangSCJLHC21,
  author       = {Ruoxi Wang and
                  Rakesh Shivanna and
                  Derek Zhiyuan Cheng and
                  Sagar Jain and
                  Dong Lin and
                  Lichan Hong and
                  Ed H. Chi},
  title        = {{DCN} {V2:} Improved Deep {\&} Cross Network and Practical Lessons
                  for Web-scale Learning to Rank Systems},
  booktitle    = {The Web Conference ({WWW})},
  pages        = {1785--1797},
  year         = {2021},
  url          = {https://doi.org/10.1145/3442381.3450078}
}

@inproceedings{DBLP:journals/corr/KingmaB14,
  author       = {Diederik P. Kingma and
                  Jimmy Ba},
  title        = {Adam: {A} Method for Stochastic Optimization},
  booktitle    = {International Conference on Learning Representations ({ICLR})},
  year         = {2015},
  url          = {http://arxiv.org/abs/1412.6980}
}

@inproceedings{DBLP:conf/nips/PaszkeGMLBCKLGA19,
  author       = {Adam Paszke and
                  Sam Gross and
                  Francisco Massa and
                  Adam Lerer and
                  James Bradbury and
                  Gregory Chanan and
                  Trevor Killeen and
                  Zeming Lin and
                  Natalia Gimelshein and
                  Luca Antiga and
                  Alban Desmaison and
                  Andreas K{\"{o}}pf and
                  Edward Z. Yang and
                  Zachary DeVito and
                  Martin Raison and
                  Alykhan Tejani and
                  Sasank Chilamkurthy and
                  Benoit Steiner and
                  Lu Fang and
                  Junjie Bai and
                  Soumith Chintala},
  title        = {PyTorch: An Imperative Style, High-Performance Deep Learning Library},
  booktitle    = {Advances in Neural Information Processing Systems (NeurIPS)},
  pages        = {8024--8035},
  year         = {2019},
  url          = {https://proceedings.neurips.cc/paper/2019/hash/bdbca288fee7f92f2bfa9f7012727740-Abstract.html}
}

@article{DBLP:journals/corr/abs-2402-03216,
  author       = {Jianlv Chen and
                  Shitao Xiao and
                  Peitian Zhang and
                  Kun Luo and
                  Defu Lian and
                  Zheng Liu},
  title        = {{BGE} M3-Embedding: Multi-Lingual, Multi-Functionality, Multi-Granularity
                  Text Embeddings Through Self-Knowledge Distillation},
  journal      = {CoRR},
  volume       = {abs/2402.03216},
  year         = {2024},
  url          = {https://doi.org/10.48550/arXiv.2402.03216}
}

@inproceedings{DBLP:conf/ijcai/GuoTYLH17,
  author       = {Huifeng Guo and
                  Ruiming Tang and
                  Yunming Ye and
                  Zhenguo Li and
                  Xiuqiang He},
  title        = {DeepFM: {A} Factorization-Machine based Neural Network for {CTR} Prediction},
  booktitle    = {International Joint Conference on
                  Artificial Intelligence ({IJCAI})},
  pages        = {1725--1731},
  year         = {2017},
  url          = {https://doi.org/10.24963/ijcai.2017/239}
}

@inproceedings{DBLP:conf/icml/RadfordKHRGASAM21,
  author       = {Alec Radford and
                  Jong Wook Kim and
                  Chris Hallacy and
                  Aditya Ramesh and
                  Gabriel Goh and
                  Sandhini Agarwal and
                  Girish Sastry and
                  Amanda Askell and
                  Pamela Mishkin and
                  Jack Clark and
                  Gretchen Krueger and
                  Ilya Sutskever},
  title        = {Learning Transferable Visual Models From Natural Language Supervision},
  booktitle    = {International Conference on Machine Learning,
                  ({ICML})},
  volume       = {139},
  pages        = {8748--8763},
  year         = {2021},
  url          = {http://proceedings.mlr.press/v139/radford21a.html}
}

\appendix
\section{Cross-Prefix Semantic Consistency of RQ-SID}
\label{app:code_consistency}

Residual codes in RQ-SID are learned from the multimodal content
space, whereas the semantics required by downstream CTR prediction
are shaped by user behavior. Consequently, increasing the
quantization depth does not necessarily provide stable additional
information and may instead introduce ambiguous semantic sharing
across different prefix paths. To examine this issue, we train a diagnostic CTR model with
hierarchically prefix-conditioned SID embeddings. For notational simplicity, we omit the item index and denote a generic
\(L\)-level code path by
\(\mathbf{c}=(c_1,c_2,\ldots,c_L)\).
Its prefix-conditioned representation is constructed as
$
\mathbf{e}^{\mathrm{SID}}
=
\sum_{l=1}^{L}
\mathbf{e}_l[c_1,\ldots,c_l].
$ For level \(l\), let
\(p=(c_1,\ldots,c_{l-1})\)
denote the preceding prefix, and let
\(\mathcal{K}^{(l)}_p\)
be the set of level-\(l\) codes observed under \(p\). We first remove the prefix-specific mean:
\begin{equation}
\boldsymbol{\mu}^{(\ell)}_p
=
\frac{1}{|\mathcal{K}^{(\ell)}_p|}
\sum_{k\in\mathcal{K}^{(\ell)}_p}
\mathbf{e}_{\ell}[p,k],
\quad
\boldsymbol{\delta}^{(\ell)}[p,k]
=
\mathbf{e}_{\ell}[p,k]
-
\boldsymbol{\mu}^{(\ell)}_p.
\end{equation}
Here, \(p=c_1\) for the second level and
\(p=(c_1,c_2)\) for the third level. For a code index \(k\) appearing under two different prefixes
\(a\) and \(b\), we measure its cross-prefix consistency as
\begin{equation}
S_{\mathrm{same}}^{(\ell)}
=
\mathbb{E}_{k,a\neq b}
\left[
\cos\left(
\boldsymbol{\delta}^{(\ell)}[a,k],
\boldsymbol{\delta}^{(\ell)}[b,k]
\right)
\right].
\end{equation}
To construct a matched random baseline, we retain the same prefix
pair \((a,b)\) but replace \(k\) under prefix \(b\) with an unrelated
valid code \(r\neq k\):
\begin{equation}
S_{\mathrm{rand}}^{(\ell)}
=
\mathbb{E}_{k,a\neq b,r\neq k}
\left[
\cos\left(
\boldsymbol{\delta}^{(\ell)}[a,k],
\boldsymbol{\delta}^{(\ell)}[b,r]
\right)
\right].
\end{equation}
For each matched comparison, we additionally compute the difference
between the same-code and random-code similarities. We report the mean and median of
mean \(\Delta\), together with the WinRate, defined as the proportion
of comparisons in which the same-code similarity exceeds the random
baseline.

\begin{table}[t]
\centering
\caption{Cross-prefix consistency of residual SID codes.}
\label{tab:sid_consistency}
\begin{tabular}{lccccc}
\toprule
Level
& \(S_{\mathrm{same}}\)
& \(S_{\mathrm{rand}}\)
& \(\Delta\)
& WinRate
& Median \(\Delta\) \\
\midrule
L2 & 0.0102 & 0.0025 & 0.0078  & 0.5005 & 0.0016 \\
L3 & 0.0293 & 0.0309 & -0.0016 & 0.4992 & 0.0000 \\
\bottomrule
\end{tabular}
\end{table}

As shown in Table~\ref{tab:sid_consistency}, the second-level codes
achieve a slightly higher average similarity than the random baseline.
However, the WinRate is nearly \(0.5\), and the median improvement is
close to zero, indicating that the apparent gain is weak and
inconsistent across code--prefix pairs. At the third level, \(S_{\mathrm{same}}\) is even lower than
\(S_{\mathrm{rand}}\), with a below-random WinRate and a zero median
improvement. Although embedding sparsity and update frequency differ
across quantization levels, these results suggest that deeper residual code semantics are highly
dependent on their preceding prefix paths. Treating globally shared deeper-level RQ-SID indices as reusable
semantic units may therefore result in semantically misaligned or
unstable knowledge sharing during downstream CTR training.

\appendix

\end{document}